\documentclass[a4paper]{article}

\usepackage{graphicx}
\usepackage{psfrag}
\usepackage{amsmath}
\usepackage{float}

\renewcommand{\theequation}{\arabic{equation}}

\begin{document}

\renewcommand{\theequation}{\arabic{equation}}

\begin{center}
{\Large {\bf $Z_N$ Gauge Theories on a Lattice and 
Quantum Memory }}

\vskip 1cm
{\Large Gaku Arakawa\footnote{e-mail address:
e101608@phys.kyy.nitech.ac.jp} and Ikuo Ichinose\footnote{e-mail 
 address: ikuo@ks.kyy.nitech.ac.jp}}  

\vskip 0.5cm

{Department of Applied Physics, Graduate School of Engineering,
Nagoya Institute of Technology,
Nagoya, 466-8555 Japan}
\end{center}
\begin{center} 
\begin{bf}
Abstract
\end{bf}
\end{center}
In the present paper we shall study $(2+1)$ dimensional
$Z_N$ gauge theories on a lattice.
It is shown that the gauge theories have two phases,
one is a Higgs phase and the other is a confinement phase.
We investigate low-energy excitation modes in the Higgs phase
and clarify relationship between the $Z_N$ gauge theories and 
Kitaev's model for quantum memory and quantum computations.
Then we study effects of random gauge couplings(RGC) which are identified
with noise and errors in quantum computations by Kitaev's model.
By using a duality transformation, it is shown that time-independent
RGC give no significant effects on the phase structure and the stability of
quantum memory and computations.
Then by using the replica methods, we study
$Z_N$ gauge theories with time-dependent RGC and show that nontrivial
phase transitions occur by the RGC.

\setcounter{footnote}{0}
\section{Introduction}
In the last few years, discrete gauge theories have got renewed
interests as a possible device for the quantum computations,
a quantum computer.
This idea was first proposed by Kitaev in his seminal paper\cite{kitaev},
and after that there appeared interesting works on this
idea\cite{preskill,dennis,wang,mochon}.
One of the most difficult problem of making a quantum computer
and performing quantum computations fault-tolerantly is the 
stability of the quantum states which participate in quantum
memory and computations.
There must be a (large) energy gap between these states and others 
in the system and also mixings of these states must be suppressed
by certain effects or selection rules.
Then one can conceive that topological interactions such as 
the Aharonov-Bohm(AB) effect may play an important role there.
The AB effect in the two spatial dimensions gives nontrivial
statistics to particles with gauge interactions, i.e., anyons.
The groundstates of the anyons are degenerate if the space is a torus
and almost no mixing occurs between them because of the
{\em topological} quantum number.
Whereas the gauge symmetry should be descrete in order to
avoid long-range interactions besides the topological ones.
Kitaev's model is based on the anyonic excitations in qubits
system.
However its detailed relationship to the gauge theory is 
still missing.

In this paper, we shall study discrete $Z_N$ gauge theories in 
$(2+1)$ dimensions.
There are two phases in these model, one is a confinement phase
and the other is a Higgs phase.
We show that Kitaev's system of qubits corresponds to
some limit of the Higgs phase of the $Z_2$ gauge models.
Stability of Kitaev's model is closely related with the
confinement-Higgs phase transition of the $Z_N$ gauge models.

This paper is organized as follows.
In Sec.2, we study $Z_N$ gauge theories which appear as a result of
spontaneous breakdown of $U(1)$ gauge symmetry.
We clarify the relationship between the gauge system and Kitaev's model
for quantum memory and computations.
In Sec.3, low-energy excitations in the Higgs phase are investigated.
There appear anyonic excitations, magnetic vortices and dyons in a
natural way as in the spontaneously broken gauge
systems in the continuum space\cite{DGT}.
In Sec.4, phase structure and effects of the (static) 
random gauge couplings(RGC)
are investigated by using a duality transformation.
The $Z_N$ gauge systems are transformed to spin systems which are
more tractable than the gauge systems.
In Sec.5, effects of the time-dependent RGC are studied by the replica
methods.
It is found that nontrivial phase transitions occur as the RGC
varies.
Section 6 is devoted to conclusion.

\section{$U(1)$ and $Z_N$ gauge theories}
Let us start with the following $U(1)$ Abelian gauge-Higgs model on a
2-dimensional($2D$) square lattice.
Hamiltonian is given by,
\begin{eqnarray}
H_{U(1)}&=&g^2\sum_{link}E^2_{xi}-{1\over g^2}\sum_{plaquette}UUUU
+{1\over \kappa}\sum_x (\Pi^\phi_x)^2
-\kappa\sum_{link}\phi^\dagger_{x+i}U^N_{xi}\phi_x-
\gamma\sum_{link}\psi^\dagger_{x+i} U^q_{xi}\psi_x  \nonumber  \\
&& +M\sum_x\psi^\dagger_x\psi_x
-\gamma\sum_{link}\varphi^\dagger_{x+i} U^{-q}_{xi}\varphi_x  
 +M\sum_x\varphi^\dagger_x\varphi_x
+\mbox{H.c.}, 
\label{HU1}
\end{eqnarray}
where $U_{xi}$ is the $U(1)$ gauge field on the link 
$(x,i)(x=\mbox{site},i=\hat{1}\mbox{ or }\hat{2})$
and $E_{xi}$ is the conjugate electric field.
The Higgs field $\phi_x \in$ $U(1)$ carries $U(1)$ charge
$N$ whereas the charge of the fermion field\footnote{We often call
$\psi_x$ and $\varphi_x$ fermion because they satisfy fermionic
anticommutation relations. As a result of the gauge interactions,
they obey anyonic statistics in the Higgs phase.
See later discussion.}$\psi_x$($\varphi_x$) 
is $q$($-q$) which is an integer.
$\Pi^\phi_x$ is the conjugate field of $\phi_x$, the
gauge coupling is $g$ and the fermion mass is $M$.
Other notations are standard.
We are interested in the case $N\neq 1$.
In this case there are two phases in the model, one is the Higgs
phase and the other is the confinement phase.
In particular in the limit $g^2 \rightarrow 0$,
the gauge field $U_{xi}$ is restricted to the pure-gauge
configuration and the model reduces to a Hamiltonian
description of the classical $3D$ XY spin model plus
the free fermion system. 
The classical $3D$ XY model exhibits a phase transition
from the magnetized phase to the disordered phase at a critical
coupling $\kappa_c$.
On the other hand for large $\kappa$, quantum fluctuations of $\phi_x$
are suppressed and low-energy excitations of the gauge and Higgs 
fields are restricted as
\begin{equation}
U^N_{xi} \sim 1, \;\;\; \phi_x \sim 1,
\label{ZNconf}
\end{equation}
up to (time-independent) local gauge transfomation.
Then we can put 
\begin{equation}
U_{xi} \sim Z_{xi},
\end{equation}
where the $Z_N$ gauge operator $Z_{xi}$ is explicitly given as
follows by $(N\times N)$ matrix,
\begin{equation}
Z_{xi}=
 \left(
    \begin{array}{ccccc}
    1&0&\cdot&\cdot&0  \\
    0&e^{{2\pi\over N}i}&0&\cdot&0  \\
    \cdot&\cdot&\cdot&\cdot&\cdot  \\
    \cdot&\cdot&\cdot&\cdot&0    \\
    0&\cdot&\cdot&\cdot&e^{{2\pi(N-1)\over N}i}
    \end{array}
\right).
\end{equation}
Corresponding to the above representation of $Z_{xi}$,
we introduce ``conjugate matrix" $X_{xi}$ as follows,
\begin{equation}
X_{xi}=
\left(
  \begin{array}{ccccc}
  0&1&0&\cdot&0 \\
  0&0&1&0&0  \\
  \cdot&\cdot&\cdot&\cdot&\cdot \\
  0&0&\cdot&0&1 \\
  1&0&\cdot&\cdot&0
  \end{array}
\right).
\end{equation}
One can easily verify the following commutation relations,
\begin{equation}
X_{xi}Z_{xi}=e^{{2\pi \over N}i}Z_{xi}X_{xi},\;\;\;
X_{xi}Z_{yj}=Z_{yj}X_{xi}\;\;
\mbox{for} \; (x,i)\neq (y,j).
\end{equation}
The electric term in Eq.(\ref{HU1}) is reduced to the following term
in the reduced $Z_N$ gauge-field space,
$$
E^2_{xi} \sim -(X_{xi}+X^\dagger_{xi}).
$$
The above result can be shown by using the eigenstates of the
electric fields as basis vectors.
Let us define an ``empty state" $|0\rangle$ as
\begin{equation}
E|0\rangle =0,
\label{E}
\end{equation}
where we have omitted link index for notational simplicity.
By using the following commutation relation,
\begin{equation}
[E,U]=U,
\label{CCR}
\end{equation}
we can show
\begin{equation}
E^2U|0\rangle=U|0\rangle \equiv |1\rangle.
\label{U1}
\end{equation}
Then the gauge field $U$ is the rising operator of the
electric field.

For the $Z_N$ case, we also define ``empty" state
for the $X$ operator,
\begin{equation}
X|0\rangle_X=|0\rangle_X, \;\;\; X^\dagger |0\rangle_X=|0\rangle_X.
\label{Xemp}
\end{equation}
The state $|0\rangle_X$ can be expressed by the eigenstates
of the $Z$ operator 
$|k\rangle_Z$, $Z|k\rangle_Z=e^{i{2\pi \over N}k}|k\rangle_Z$,
\begin{equation}
|0\rangle_X={1\over \sqrt{N}}\Big(|1\rangle_Z+|2\rangle_Z+\cdots
+|N\rangle_Z\Big).
\label{0X}
\end{equation}
Then one can easily show,
\begin{equation}
Z|0\rangle_X=|1\rangle_Z +e^{{2\pi \over N}i}|2\rangle_Z+\cdots
+e^{2\pi{N-1\over N}i}|N\rangle_Z,
\end{equation}
and therefore
\begin{eqnarray}
XZ|0\rangle_X&=&e^{{2\pi \over N}i}Z|0\rangle_X,  \\
\Big(X+X^\dagger\Big)Z|0\rangle_X &=& 2\cos \Big({2\pi \over N}\Big)
\; Z|0\rangle_X.
\label{X+X}
\end{eqnarray}
From Eqs.(\ref{Xemp}) and (\ref{X+X}), $Z$ is the lowering operator
of $X+X^\dagger$ and therefore,
\begin{equation}
E^2_{xi} \sim -(X_{xi}+X^\dagger_{xi}),
\label{EX2}
\end{equation}
up to irrelevant additive and multiplicative constants.
Then for large $\kappa$, the $U(1)$ gauge theory (\ref{HU1})
reduces to the following $Z_N$ gauge theory,
\begin{eqnarray}
H_{\cal T}&=& H_Z+H^\psi_Z+H^\varphi_Z,  \nonumber \\
H_Z &=& -\lambda_1\sum X_{xi}-\lambda_2 \sum ZZZZ + \mbox{H.c.}, \nonumber \\
H^\psi_Z &=& - \gamma \sum \psi^\dagger_{x+i} Z^q_{xi}\psi_x
+M\sum \psi^\dagger_x\psi_x+\mbox{H.c.}, \nonumber\\
H^\varphi_Z &=& - \gamma \sum \varphi^\dagger_{x+i} Z^{-q}_{xi}\varphi_x
+M\sum \varphi^\dagger_x\varphi_x+\mbox{H.c.}, 
\label{HZ}
\end{eqnarray}
where $\lambda_1$ and $\lambda_2$ are coupling constants of 
the $Z_N$ gauge theory
and they relate to the $U(1)$ gauge coupling
$g^2$ as $\lambda_1 \sim g^2$ and
$\lambda_2 \sim 1/g^2$.

The above ``derivation" of the $Z_N$ gauge theory (\ref{HZ}) from the $U(1)$
gauge system (\ref{HU1}) is rather sketchy but it might be useful
for realization of discrete gauge systems in architecture of the
quantum computers.
For example, spontaneous breaking of 
$U(1)$ gauge symmetry occurs in the superconductivity.
In most of the superconductors including the high-temperature
ones, the ``Cooper pair" carries electric charge $2e$.
Then a discrete $Z_2$ gauge system close to the present one
might be realized in some superconductors.
The Hamiltonian (\ref{HZ}) is directly obtained from the
path-integral formalism of the $Z_N$ gauge theory on 
$3D$ lattice by taking the continuum limit of the time-like
direction.
In the $3D$ $Z_N$ gauge theory, there exist two phases, i.e.,
confinement and Higgs phases as we show later on.
Phase transition occurs at a certain critical coupling 
$(\lambda_1/\lambda_2)_c$.
In the original $U(1)$ gauge theory, there exists a {\em critical 
line} connecting the XY phase transition at $(g=0,\kappa=\kappa_c)$
and $Z_N$ gauge phase transition at $(g=g_c, \kappa=\infty)$
(see Fig.1)\cite{Z2}.

Physical state of the system (\ref{HZ}) must be gauge-invariant
and this condition is given as follows,
\begin{equation}
G_x \equiv \Big(\prod_{(y,i)\in x}\tilde{X}_{yi}\Big)
e^{-{2\pi q \over N}i(\psi^\dagger_x\psi_x
-\varphi^\dagger_x\varphi_x)}, \;\; 
G_x|phys\rangle=|phys\rangle,
\label{Phys}
\end{equation}
where $(y,i) \in x$ denotes 4 links emanating from site $x$
and $\tilde{X}_{yi}=X_{yi}$ for $y=x$ whereas 
$\tilde{X}_{yi}=X^\dagger_{xi}$ for $y-i=x$.
It is proved that $G_x$ is the gauge-transformation operator at site
$x$ and the Hamiltonian $H^\psi_Z+H^\phi_Z$ in Eq.(\ref{HZ}) commutes with
$G_x$.

Recently Kitaev proposed a 2-dimensional qubits system for
fault-tolerant quantum memory and computations\cite{kitaev}.
This system is closely related to the $Z_2$ gauge theory and
contains ``anyonic excitations".
The system is defined on a torus and
the Hamiltonian is given as follows in our notation,
\begin{equation}
H_K=-\sum_x \prod_{(y,i)\in x}{X}_{yi}-\sum_{pl}ZZZZ,
\label{HK}
\end{equation}
where $Z_{xi}$ and $X_{xi}$ are explicitly given by the Pauli
matrices in the $Z_2$ case, i.e., $Z=\sigma^z$ and $X=\sigma^x$.
The groundstates and excited states of the Hamiltonian (\ref{HK})
are easily obtained since the first and second terms of (\ref{HK}) commute
with each other.
The groundstates satisfy
\begin{equation}
\prod_{(y,i)\in x}{X}_{yi}|GS\rangle_K=|GS\rangle_K, \;\;\; 
\prod_{pl}Z|GS\rangle_K=|GS\rangle_K,
\label{Grand}
\end{equation}
for all sites and plaquettes.
The groundstates are four-fold degenerate on the torus,
as we explain in the following section.
These degenerate lowest-energy states form basis for quantum 
memory\cite{kitaev}.

The first excited states are explicitly given by 
\begin{equation}
\prod_{(y,i)\in x}{X}_{yi}|1st\rangle_K=-|1st\rangle_K, \;\;\mbox{or} \;\;
\prod_{pl}Z|1st\rangle_K=-|1st\rangle_K,
\label{Excite}
\end{equation}
for some specific site or plaquette and otherwise they 
satisfy Eq.(\ref{Grand}).
It is not so difficult to see that Kitaev's model is equivalent to
the model (\ref{HZ}) with $N=2, \gamma=0, M=2, q=1$ and 
$\lambda_1=0, \lambda_2=1$.
With these parameters and the physical state condition (\ref{Phys}),
the groundstates of the gauge model are given as
\begin{equation}
\prod_{pl}Z|GS\rangle_Z=|GS\rangle_Z, \;\; 
\psi^\dagger_x\psi_x|GS\rangle_Z=0, \;\; 
\varphi^\dagger_x\varphi_x|GS\rangle_Z=0,
\label{GS2}
\end{equation}
for all plaquettes and sites.
From (\ref{Phys}), the second and third conditions of (\ref{GS2}) mean
$\prod_{(y,i)\in x}{X}_{yi}|GS\rangle_Z=|GS\rangle_Z$.
On the other hand, the first excited states of the gauge
system are given by,
\begin{equation}
\prod_{pl}Z|1st\rangle^V_Z=-|1st\rangle^V_Z, \;\;\mbox{or} \;\;
\psi^\dagger_x\psi_x|1st\rangle^\psi_Z=|1st\rangle^\psi_Z, \;\;\mbox{or} \;\;
\varphi^\dagger_x\varphi_x|1st\rangle^\varphi_Z=|1st\rangle^\varphi_Z,
\label{Excite2}
\end{equation}
for some specific plaquette or site.
From Eq.(\ref{Phys}), the second condition in (\ref{Excite2}) is
equivalent to 
$\prod_{(y,i)\in x}{X}_{yi}|1st\rangle^\psi_Z=-|1st\rangle^\psi_Z$
for $e^{\pi i \psi^\dagger_x\psi_x}|1st\rangle^\psi_Z=-|1st\rangle^\psi_Z$
and energy increases by $2$ because of the mass term 
in (\ref{HZ}) with $M=2$.
Similarly for the other fermion $\varphi_x$.
In the original paper by Kitaev\cite{kitaev}, relationship between
his model and gauge theories was slightly discussed but full
relationship was missing.
In the following sections we shall study phase structure of the 
present $Z_N$ gauge model, low-energy excitations, effects of
random gauge couplings, etc.
All these discussions give an important insight to the stability
problem of Kitaev's model.

\section{Low-energy excitations in the Higgs phase}
As we show in the following section, there are two phases in the 
$Z_N$ gauge theory in $(2+1)$ dimensions $H_Z$ in (\ref{HZ}).
For large $\lambda_2/\lambda_1$, fluctuation of the gauge field $Z_{xi}$ 
is small and the Higgs phase is realized
whereas for small $\lambda_2/\lambda_1$, the gauge field $Z_{xi}$
fluctuates strongly and the confinement phase
is realized.
The Higgs phase of the model can be used for a quantum memory.
Coupling of the matter fields $\psi_x$ etc. enhances the Higgs phase.

Let us study the model on the {\em torus} and 
focus on the Higgs phase for large $\lambda_2/\lambda_1$.
In particular for $\lambda_1=0$, the groundstates are given by
Eq.(\ref{GS2}) and low-energy excited states are {\em particle
states} of $\psi_x$, $\varphi_x$ and
states of {\em plaquette magnetic excitation}
or {\em vortex}, i.e.,
$\prod_{pl}Z|1st\rangle^V_Z=-|1st\rangle^V_Z$ for specific plaquette.
As we study the model on the torus, we have the following ``trivial" identities
\begin{equation}
\prod_{\mbox{\small all sites}}\prod_{(y,i)\in x}\tilde{X}_{yi}=1, \;\;\;
\prod_{\mbox{\small all pl's}}\prod_{pl}Z=1,
\label{id}
\end{equation}
and therefore the above excitations must appear in pairs.
As the groundstates satisfy Eq.(\ref{GS2}), there is no
mangetic flux in each plaquette.
Then one may think that the groundstate is unique.
However this is not the case.
There are two nontrivial cycles on the torus, and let us call them
a-cycle and b-cycle, i.e., noncontractible closed paths.
We introduce the dual lattice in the usual way, and choose certain 
noncontractible closed loops on the original and dual lattices.
We use notations such that $C^a_Z$($C^b_Z$) for a suitably chosen
closed loop corresponding to 
the a-cycle(b-cycle) on the {\em original} lattice and $C^a_X$($C^b_X$) for 
a loop corresponding to the a-cycle(b-cycle) on the {\em dual} lattice. 
Later discussion does not depend on the choice of the loops.
Then we define the following operators, $Z_a,Z_b, X_a$ and $X_b$,
\begin{eqnarray}
&& Z_a=\prod_{C^a_Z}Z_{xi}, \; \; 
  Z_b=\prod_{C^b_Z}Z_{xi}, \nonumber \\
&&  X_a=\prod_{C^a_X}X_{xi}, \; \; 
  X_b=\prod_{C^b_X}X_{xi},
\label{ZX}
\end{eqnarray}
where $X_{xi}$'s in $X_a$ cross $C^a_X$ and similarly for $X_b$.
These operators are obviously invariant under gauge transformation
and commute with $H_Z$ when $\lambda_1=0$.
Furthermore they satisfy the following commutation relations,
\begin{equation}
X_aZ_b=e^{{2\pi \over N}i}Z_bX_a, \;\; X_bZ_a=e^{{2\pi \over N}i}Z_aX_b,
\label{CR}
\end{equation}
and otherwise commute.
Therefore the groundstates are eigenstate of the the operators, e.g., 
$Z_a$ and $Z_b$ and they are $N^2$-fold degenerate.
This result holds even in the presence of the fermions $\psi_x$
and $\varphi_x$ since $Z_a$ and $Z_b$ commute with $H_{\cal T}$
for vanishing $\lambda_1$ in Eq.(\ref{HZ}).

Fermions $\psi_x$ and $\varphi_x$ move in an unfluctuating ``background"
field of $Z_{xi}$'s with vanishing magnetic field.
However they distinguish the above $N^2$-fold degenerate $Z$'s groundstates.
In fact while $\psi_x$(or $\varphi_x$) fermion moves along 
a closed loop of the a-cycle,
it acquires phase factor which is an eigenvalue of $(Z_a)^q$,
and similarly for the b-cycle.
Then the Higgs phase is a ``topologically ordered" phase.
The $N^2$ groundstates work as {\em qudit} for quantum memory
and the quantum states of the qudit are distinguishable by using
matter fields like $\psi_x$.

Let us discuss excitations in detail.
As we explained above, the fermions must appear in a pair.
Two-fermion state at sites $x$ and $y$ is explicitly given as,
\begin{equation}
|F;C_{xy}\rangle =
\psi^\dagger_y\Big(\prod_{C_{xy}}Z^q\Big)\varphi^\dagger_x|GS\rangle_Z,
\label{Fstate}
\end{equation}
where $C_{xy}$ is a certain path on the {\em original} lattice
connecting $x$ and $y$,
and the state (\ref{Fstate}) obviously satisfies the physical-state
condition (\ref{Phys}).
On the other hand two-vortex state at {\em dual} sites $x^\ast$ and 
$y^\ast$ is given as,
\begin{equation}
|V;\tilde{C}_{x^\ast y^\ast}\rangle  
 = \Big(\prod_{\tilde{C}_{x^\ast y^\ast}}X\Big)|GS\rangle_Z,
\label{Vstate}
\end{equation}
where $\tilde{C}_{x^\ast y^\ast}$ is a certain path on the {\em dual} lattice
connecting $x^\ast$ and $y^\ast$ and $X$'s in (\ref{Vstate})
are on the links crossing $\tilde{C}_{x^\ast y^\ast}$(see Fig.2).
This state is also a physical state.
Other physical excitations are produced by appling the gauge-invariant
operators in Eqs.(\ref{Fstate}) and (\ref{Vstate}) succsessively on the
groundstates.

Fermionic excitations and magnetic vortices satisfy a nontrivial
statistics.
This is an Aharonov-Bohm effect of the $Z_N$ gauge theory.
To see this, we consider the state like
\begin{equation}
\psi^\dagger_{y_1}\Big(\prod_{C_{x_1y_1}}Z^q\Big)\varphi^\dagger_{x_1}\cdot
\Big(\prod_{\tilde{C}_{x_2^\ast y_2^\ast}}X\Big)
|GS\rangle_Z,
\label{FV}
\end{equation}
and assume that the paths $C_{x_1y_1}$ and 
$\tilde{C}_{x_2^\ast y_2^\ast}$ do not entangle with each other.
Let us move the $\varphi_x$ fermion at $x_1$ around the vortex at
$x_2^\ast$ once counterclockwise (and not $y_2^\ast$)
and then return it to the original position $x_1$.
The resultant path $C'_{x_1y_1}$ encircles $x_2^\ast$ once
and $C'_{x_1y_1}$ and $\tilde{C}_{x_2^\ast y_2^\ast}$
cross with each other.
Then the state can be written as 
\begin{equation}
\psi^\dagger_{y_1}\Big(\prod_{C'_{x_1y_1}}Z^q\Big)\varphi^\dagger_{x_1}\cdot
\Big(\prod_{\tilde{C}_{x_2^\ast y_2^\ast}}X\Big)
|GS\rangle_Z=
\psi^\dagger_{y_1}\Big(\prod_{C_{x_1y_1}}Z^q\cdot
\prod_{C_{\mbox{\footnotesize closed}}}Z^q\Big)\varphi^\dagger_{x_1}\cdot
\Big(\prod_{\tilde{C}_{x_2^\ast y_2^\ast}}X\Big)|GS\rangle_Z,
\label{FV2}
\end{equation}
where $C_{\mbox{\footnotesize closed}}$ is the closed path 
$(C'_{x_1y_1} \cdot C^{-1}_{x_1y_1})$ which
encircles $x_2^\ast$ once and has a single common link 
(or odd number of links) with $\tilde{C}_{x_2^\ast y_2^\ast}$.
Because of the nontrivial commutation relation between
$Z_{xi}$ and $X_{xi}$ and (\ref{GS2}), the resultant state differs from the
original one by the phase factor $e^{{2q\pi \over N}i}$,
\begin{equation}
\psi^\dagger_{y_1}\Big(\prod_{C_{x_1y_1}}Z^q\cdot
\prod_{C_{\mbox{\footnotesize closed}}}Z^q\Big)\varphi^\dagger_{x_1}\cdot
\Big(\prod_{\tilde{C}_{x_2^\ast y_2^\ast}}X\Big)|GS\rangle_Z
=e^{{2q\pi \over N}i}
\psi^\dagger_{y_1}\Big(\prod_{C_{x_1y_1}}Z^q\Big)\varphi^\dagger_{x_1}\cdot
\Big(\prod_{\tilde{C}_{x_2^\ast y_2^\ast}}X\Big)
|GS\rangle_Z.
\end{equation}

The above anyonic properties of the low-energy excitations are
closely related with the groundstate degeneracy.
In the continuum spacetime, a Chern-Simons(CS) gauge theory
is often employed for describing anyons which are a nontrivial 
representation of the braid group.
In anyon systems on a torus, movement of an anyon along noncontractible
loops like the a-cycle and/or b-cycle is a nontrivial element of the
braid group.
On the torus, the zero modes of the CS gauge field play an important role 
and the groundstate wave function of anyons becomes multi-component because
of the zero modes\cite{IO}.
Similar phenomenon occurs in the present $Z_N$-gauge system
as we explained above.

One may conceive that the system has dyonic excitations
as in the continuum theories\cite{DGT}.
The answer is positive.
Dyon $d_x$ with ``electric charge" $Q_E$ and ``magnetic charge" $R$
is described by the following Hamiltonian,
\begin{equation}
H_D=-\sum d^\dagger_{x+i}Z^{Q_E}_{xi}X^R_{\bar{xi}}d_x+\mbox{H.c.},
\label{HD}
\end{equation}
where we assume that the fields $d_x$ and $d^\dagger_x$
themselves satisfy the fermionic commutation relations
for simplicity.
The link $(\bar{xi})$ is associated with the link $(xi)$
and defined as follows,
\begin{equation}
\mbox{link}\; (\bar{xi})=\left\{ 
      \begin{array}{ll}
      (x+1,2)& \;\; \mbox{for $i=1$},  \\
      (x+2,1)& \;\; \mbox{for $i=2$}.
      \end{array}
     \right.
\label{link}
\end{equation}
From the above definition (\ref{HD}) and (\ref{link}), it is obvious that 
the electric charge $Q_E$ of the dyon $d_x$is located at the site $x$ whereas
its magnetic charge $R$ is located at the nearest-neighbor plaquette
(see Fig.3). 
Regularization is naturally introduced by the spatial lattice.
It is not so difficult to show that the above dyon satisfies
nontrivial representation of the braid group and there appears
the phase factor like $-\exp (\pm {2(Q_E+R)\pi i \over N})$ when two
dyons interchange with each other.

When we turn on the parameter $\lambda_1$ in $H_Z$ (\ref{HZ}),
the operator $Z_a$ and $Z_b$ do {\em not} commute with $H_{\cal T}$
anymore and therefore degeneracy of the groundstate disappears.
This stems from the fact that because of the term $\lambda_1\sum X_{xi}$, 
$Z_{xi}$ becomes dynamical and it fluctuates 
quantum mechanically and then genuine anyonic properties of the
low-energy excitations break down.
However for small $\lambda_1$, there is still an energy gap between the
$N^2$ ``groundstates" with fine structure and the other excited states.
Furthermore, these $N^2$ states are far apart with each other in the
quantum-mechanical configuration space and are hardly mixed if
the torus is sufficiently large.
Therefore the system with small value of $\lambda_1$ is still suited for
a quantum memory as Kitaev suggested first.
However as $\lambda_1$ increases, a phase transition occurs as 
we show in the following section.
In the new phase, a confinement phase, the gauge field fluctuates
randomly and the system is useless
as a quantum memory.\footnote{It is very interesting to see that similar
gauge-theory argument can be applied to neural network models
for brain\cite{matsui}. There Higgs phase corresponds to good
brains and the confinement phase to dementia.}

\section{Duality transformation, phase transition and random
gauge couplings}

In the previous section, we discussed that for the quantum
memory and commputations the Higgs phase must be realized 
in the present system.
In this section we shall study the phase structure of the 
gauge-theory model $H_Z$ in (\ref{HZ}).
To this end, the system is defined on a large spatial square
lattice.
We shall perform a duality transformation which transforms
the gauge-theory model into a more tractable spin model.
For the $Z_2$ gauge theory, the duality transfomation is
discussed in Kogut's review article\cite{kogut}.

Let us consider the pure gauge system $H_Z$ in (\ref{HZ})
with the physical state condition,
\begin{equation}
\prod_{(y,i)\in x}\tilde{X}_{yi}=X_{x1}X_{x,-1}X_{x2}X_{x,-2}
=X_{x1}X^\dagger_{x-1,1}X_{x2}X^\dagger_{x-2,2}=1.
\label{cond2}
\end{equation}
By solving the above condition (\ref{cond2}), the operator $X_{x2}$
is given as follows by the remaining operators,
\begin{equation}
X_{x2}=X^\dagger_{x1}X^\dagger_{x,-1}X^\dagger_{x-2,1}X^\dagger_{x-2,-1}\cdots.
\label{X2}
\end{equation}
As the ``conjugate" operators $Z_{x2}$ of $X_{x2}$ commute 
with the Hamiltonian $H_Z$, we can set it as a constant, $Z_{x2}=1$.

Then we introduce the following dual operators $W_{x^\ast}$ and 
$V_{x^\ast}$ which reside on sites of the {\em dual} lattice,
\begin{eqnarray}
&& W_{x^\ast}=\prod_{(yi)\in x^\ast}Z_{yi},  \nonumber  \\
&& V_{x^\ast}=\prod_{l\ge 0}X_{x-2l,1},
\label{WV}
\end{eqnarray}
where $(yi) \in x^\ast$ denotes $4$ links to the plaqutte on the
original lattice which is dual to the site $x^\ast$ of the dual
lattice.
From the definition (\ref{WV}), one can easily verify 
relations like,
\begin{eqnarray}
&& W^N_{x^\ast}=V^N_{x^\ast}=1, \;\; 
V_{x^\ast} W_{x^\ast}= e^{i{2\pi \over N}} W_{x^\ast} V_{x^\ast}, \nonumber \\
&& V_{x^\ast} W_{y^\ast}= W_{y^\ast} V_{x^\ast}, \;\; 
\mbox{for $x^\ast\neq y^\ast$}, \nonumber \\
&& V_{x^\ast} W^\dagger_{x^\ast}= e^{-i{2\pi \over N}} W^\dagger_{x^\ast}
 V_{x^\ast},
\label{WV2}
\end{eqnarray}
\begin{equation}
V_{x^\ast}V^\dagger_{x^\ast-2}=X_{x1}, \;\; 
V^\dagger_{x^\ast}V_{x^\ast-1}=X_{x2}.
\label{VVX}
\end{equation}
From Eqs.(\ref{WV2}) and (\ref{VVX}), the Hamiltonian $H_Z$ in (\ref{HZ})
can be rewritten in terms of $V_{x^\ast}$ and $W_{x^\ast}$,
\begin{equation}
H_Z=-\lambda_1 \sum_{x^\ast, i=1,2} V_{x^\ast}V^\dagger_{x^\ast-i}
-\lambda_2 \sum_{x^\ast}W_{x^\ast}+\mbox{H.c.}
\label{HZ2}
\end{equation}
The above quantum Hamiltonian (\ref{HZ2}) is nothing but that of the $3D$
classical $Z_N$ Ising model(the clock model) which is obtained 
by the transfer-matrix methods and taking the continuum limit of
one direction.

The Hamiltonian (\ref{HZ2}) is more tractable than the original
one (\ref{HZ}).
There are two phases, i.e., ordered  and disordered phases,
and a phase transition occurs as the value $\lambda_1/\lambda_2$
varies.
For small $\lambda_1/\lambda_2$ limit, the groundstate is given by
\begin{equation}
W_{x^\ast}|0\rangle_S =W^\dagger_{x^\ast}|0\rangle_S=|0\rangle_S.
\label{spinGS}
\end{equation}
In the representation,
\begin{equation}
W_{x^\ast}=
 \left(
    \begin{array}{ccccc}
    1&0&\cdot&\cdot&0  \\
    0&e^{{2\pi\over N}i}&0&\cdot&0  \\
    \cdot&\cdot&\cdot&\cdot&\cdot  \\
    \cdot&\cdot&\cdot&\cdot&0    \\
    0&\cdot&\cdot&\cdot&e^{{2\pi(N-1)\over N}i}
    \end{array}
\right),
\end{equation}
the above groundstate $|0\rangle_S$ is explicitly given as,
\begin{equation}
|0\rangle_S=\prod_{x^\ast} |0\rangle_{x^\ast}, \;\; 
|0\rangle_{x^\ast}=
\left(
   \begin{array}{c}
   1 \\
   0 \\
   \cdot \\
   \cdot \\
   0
  \end{array}
\right).
\end{equation}
For small but nonvanishing $\lambda_1/\lambda_2$, the groundstate
is obtained by the usual perturbative calculation,
and the term $V_{x^\ast}V^\dagger_{x^\ast-i}$ tilts nearest-neighbor 
$W_{x^\ast}$ and $W_{x^\ast-i}$ by $\pm {2\pi \over N}$, respectively.
In this phase, there is no ``magnetization", i.e., 
\begin{equation}
_S\langle 0|V_{x^\ast}|0\rangle_S=0.
\label{mag}
\end{equation}
Low-energy excitations are given by,
\begin{equation}
{1\over \sqrt{N_s}}\sum_{x^\ast}e^{ip\cdot x^\ast}V_{x^\ast}
|0\rangle_S, \;\;  
{1\over \sqrt{N_s}}\sum_{x^\ast}e^{ip\cdot x^\ast}V^\dagger_{x^\ast}
|0\rangle_S,
\label{excite}
\end{equation}
where 2-vector $p$ is a momentum and $N_s$ is the number of the sites.
Excited energy of the above states (\ref{excite}) can be easily 
calculated and obtained as follows for small $\lambda_1/\lambda_2$,
\begin{equation}
E=2\lambda_2\Big(1-\cos ({2\pi \over N})\Big)+\cdots.
\label{Energy}
\end{equation}
From (\ref{Energy}), the energy gap is a decreasing function of $N$.

For large $\lambda_1/\lambda_2$, on the other hand, 
the groundstate of the spin system (\ref{HZ2}) is given by 
\begin{equation}
V_{x^\ast}|\tilde{0}\rangle_S=e^{i\alpha}|\tilde{0}\rangle_S, \;\;
V^\dagger_{x^\ast}|\tilde{0}\rangle_S=e^{-i\alpha}|\tilde{0}\rangle_S,
\label{spinGS2}
\end{equation}
where $e^{i\alpha} \in Z_N$ and therefore it is $N$-fold
degenerate.\footnote{This result of the spin system does
{\em not} mean that the original gauge system has the degenerate
groundstates. Actually from (\ref{HZ}), the groundstate satisfies
$X_{xi}|GS; \mbox{gauge}\rangle=1$ for all links $(xi)$ in the 
large $\lambda_1/\lambda_2$ limit.}
There is a nonvanishing magnetization for large $\lambda_1/\lambda_2$, 
\begin{equation}
_S\langle\tilde{0}|
V_{x^\ast}|\tilde{0}\rangle_S \neq 0.
\label{mag2}
\end{equation}
From Eqs.(\ref{WV}), (\ref{mag2}) and the discussion in the previous section,
it is obvious that {\em vortex condensation} occurs
in the gauge-system state corresponding to $|\tilde{0}\rangle_S$.
This means that for large $\lambda_1/\lambda_2$ the confinement
phase is realized and therefore the gauge system does not work as 
a quantum memory.
This result is important for the architecture of the quantum computer.

It is interesting and also important to study another type of
disturbance for realization of the Higgs phase, or, a good quantum
memory, i.e., the effect of random gauge couplings which corresponds to
noise and errors in quantum computations.
In this section, we consider {\em static} random gauge coupling(RGC) with
random variables $\tau_{x^\ast} \in Z_N$, and the Hamiltonian
is given by,
\begin{equation}
H^R_Z=-\lambda_1\sum_{link} X_{xi}-\lambda_2\sum_{pl}
\tau_{x^\ast} ZZZZ.
\label{HRZ}
\end{equation}
We assume a simple local correlation for the random variables,
\begin{equation}
[\tau_{x^\ast}\tau_{y^\ast}]\propto \delta_{x^\ast y^\ast},
\label{CR2}
\end{equation}
where $[\cdots]$ denotes the ensemble average.

It seems rather difficult to study the above random gauge system
(\ref{HRZ}).
However by using the duality transformation (\ref{WV}),
we can rewrite the Hamiltonian $H^R_Z$ as in the nonrandom case
which we studied above,
\begin{equation}
H_Z=-\lambda_1 \sum_{x^\ast, i=1,2} V_{x^\ast}V^\dagger_{x^\ast-i}
-\lambda_2 \sum_{x^\ast}\tau_{x^\ast}W_{x^\ast}+\mbox{H.c.}
\label{HRZ2}
\end{equation}
Then we {\em redefine} the dual operators $W_{x^\ast}$ as follows,
\begin{equation}
\tilde{W}_{x^\ast}\equiv \tau_{x^\ast}W_{x^\ast}.
\label{W2}
\end{equation}
One can easily verify that the new operator $\tilde{W}_{x^\ast}$
and the old one $W_{x^\ast}$
satisfy exactly the same operator equations in (\ref{WV2}),
and also there are
{\em no} spatial correlations of $\tilde{W}_{x^\ast}$ because of 
(\ref{CR2}).
Then $H^R_Z$ is equivalent to the original $H_Z$ and the random
system has the same phase structure with the nonrandom one.
The groundstate, excitated states, etc. are different in the two
systems but there exists one-to-one correspondence between them.
This result can be partly seen in the original gauge system (\ref{HRZ}).
For {\em static} RGC $\tau_{x^\ast}$ and $\lambda_1=0$, 
there is a {\em unique} $Z$-field configuration of the lowest energy 
up to local gauge transformations.
Vortex excitations are generated by applying the string operator 
$(\prod X_{xi})$ on this lowest-energy configuration as in the
nonrandom case.
Then there is one-to-one correspondence.
The perturbative term $\lambda_1 X_{xi}$ generates a pair of
vortices in a nearest-neighbor plaquettes from the lowest-energy
state as in the nonrandom system.

In the following section, we shall study random $Z_2$ gauge system
with full RGC by the replica methods and show that nontrivial phase
structure appears.


\section{Replica mean-field theory}

In this section we shall study the $d$-dimensional random 
$Z_2$ gauge theories by the replica mean-field theory(RMFT).
RMFT has been often applied to the random spin systems and spin-glass
problems.
The {\em random gauge theories} have been less studied and
as far as we know there is no systematic studies on
the random gauge theories by the replica methods.
Numerical Monte-Carlo simulations are also useful to obtain 
phase diagram in the $p-T$ plane, where $p$ is the concentration
of plaquettes of ``wrong sign" and $T$ is the temperature\cite{OAIM,wang2}.
In the previous sections we used the Hamiltonian formalism,
but in this section we employ the path-integral Lagrangian
formalism since the path-integral formalism is more
suitable for the MFT.

Let us first study the nonrandom $Z_2$ gauge theory on a $d$-dimensional
lattice by the MFT\cite{drouffe}.
The partition function $Z$ is given by,
\begin{equation}
Z=\mbox{Tr} \; e^{-\beta S}, \; \; 
S=-\sum_{pl}\sigma\sigma\sigma\sigma-h\sum_{link}\sigma_{xi},
\label{Z}
\end{equation}
where the $Z_2$ gauge variables $\sigma_{xi}$($i=1,\cdots,d$) take $\pm 1$, 
Tr means $\sum_{\sigma_{xi}=\pm 1}$, $\beta$ is inverse temperature
and $h$ is an external ``magnetic field".
It is not so difficult to drive MFT action $S_M$.
To this end, let us decompose $\sigma_{xi}$ as 
$\sigma_{xi}=U_0+\delta \sigma_{xi}$
where $U_0$ is the MF for $\sigma_{xi}$ and $\delta \sigma_{xi}$
is the fluctuation from it\cite{elitzur}.
In terms of the new variables,
\begin{eqnarray}
S&=&-U^4_0N_P-U^3_0\cdot 2(d-1)\sum_{link}\delta \sigma_{xi}-h
\sum_{link}\sigma_{xi} +O((\delta \sigma_{xi})^2) \nonumber  \\
&=&-U^4_0N_P-U^3_0\cdot 2(d-1)\sum_{link}\sigma_{xi}+2(d-1)U^4_0N_L
    -h\sum_{link}\sigma_{xi} +O((\delta \sigma_{xi})^2),
\label{MF1}
\end{eqnarray}
where $N_P$ and $N_L$ are the numbers of plaquettes and links 
of the lattice, respectively and $N_L={2\over d-1}N_P$.
From (\ref{MF1}), $S_M$ is obtained as,
\begin{equation}
S_M=3U_0^4N_P-\{2(d-1)U_0^3+h\}\sum_{link}\sigma_{xi}.
\label{SM}
\end{equation}
Then it is straightforward to calculate the partition function
from $S_M$ in (\ref{SM}),
\begin{eqnarray}
Z_{MF}&=& \mbox{Tr} e^{-\beta S_M} \nonumber  \\
&=&e^{-3\beta U_0^4N_P}\Big[2 \cosh \beta \{2(d-1)U^3_0+h\}\Big]
^{N_L}.
\label{ZMF}
\end{eqnarray}
The ``magnetization" $m$ per link is calculated from (\ref{ZMF}) as
\begin{eqnarray}
m&=& {1\over Z_{MF}N_L}{\partial Z_{MF} \over \partial h}  \nonumber  \\
&=&\tanh \beta \{2(d-1)U^3_0+h\}.
\label{m}
\end{eqnarray}
Similarly the free energy is obtained as,
\begin{equation}
F=-{1\over \beta}{2\over d-1}N_P \log \Big[2\cosh \beta\{2(d-1)U^3_0+h\}
\Big] +3U_0^4N_P.
\label{F}
\end{equation}

Numerical calculation of the free energy $F$ in (\ref{F}) is given in 
Fig.4 as a function $U_0$ for vanishing $h$ and at various inverse 
temperatures $\beta$.
From Eqs.(\ref{m}) and (\ref{F}), it is verified that the magnetization $m$
is equal to the value of  $U_0$ at stationary points of $F=F(U_0)$.
Result in Fig.4 shows that there is a first-order phase transition
as the temperature varies and at low temperature the magnetiztion $m$
is novanishing.

It is known that there is a {\em second-order} phase transition in $3D$
$Z_2$ gauge theory which is dual to the $3D$ Ising model
as we showed in the previous section\cite{kogut}.
This means that the MFT gives correct results only at large spatial
dimensions as it is well known for the spin systems, etc.
However we believe that the MFT is still useful for obtaining rough
estimations of the physical quantities, phase stucture of systems, etc.

Let us turn to random gauge theories(RGT).
We study the random $Z_2$ theory with the following action,
\begin{equation}
S_R=-\sum_{pl}J_p\sigma\sigma\sigma\sigma,
\label{RGT}
\end{equation}
where we assume that the RGC $J_p$ has the probability distribution like,
\begin{equation}
P(J_p)={1 \over J\sqrt{2\pi}}\exp 
\Big\{-{1\over 2J^2}(J_p-J_0)^2\Big\},
\label{PJ}
\end{equation}
with positive parameters $J$ and $J_0$. 
We choose (\ref{PJ}) in which $J_p$ takes continuous real value
instead of the discrete distribution $J_p=\pm J_0$,
because it is more tractable.

We apply replica tricks to the above RGT and then the partition function
is given as,
\begin{equation}
[Z^n]=\int \Big\{\prod_pdJ_pP(J_p)\big\}
\mbox{Tr} \exp \Big(\beta\sum_pJ_p\sum_{\alpha=1}^n\prod_p
\sigma^\alpha +\beta h \sum_{link} \sigma^\alpha\Big),
\label{Zn}
\end{equation}
where $\alpha$ is the replica index which takes $\alpha=1,\cdots,n$,
and we shall take the limit $n\rightarrow 0$ in the final stage
of the calculation.
Because of the replica tricks,
integration over $J_p$ can be done for each $p$,
\begin{equation}
\int dJ_p e^{-{1\over 2J^2}(J_p-J_0)^2}e^{\beta J_p
\sum_{\alpha=1}^n \prod_p \sigma_\alpha}
=e^{{1\over 2}\beta^2J^2\sum_{\alpha,\beta}\prod_p\sigma^\alpha
\prod_p\sigma^\beta
+J_0\beta\sum_\alpha \prod_p \sigma^\alpha}
\label{U}
\end{equation}
We introduce the MF $U_{0\alpha}$ for $\sigma^\alpha_{xi}$ 
and the glass MF(GMF)
$Q_{\alpha\beta}$ for $\sigma^\alpha_{xi}\sigma^\beta_{xi}$.
Then the terms in the action (\ref{U}) can be rewritten as follows
as in the MFT for the nonrandom case (\ref{SM}),
\begin{eqnarray}
&& \sum_{pl}\prod_{p}\sigma^\alpha \rightarrow
   -3N_PU^4_{0\alpha}+4CU^3_{0\alpha}
   \sum_{link}\sigma^\alpha_{xi},  \nonumber  \\
&& \sum_{pl}\prod_p\sigma^\alpha\prod_p\sigma^\beta \rightarrow
   -3N_PQ_{\alpha\beta}^4+4CQ_{\alpha\beta}^3\sum_{link}
  \sigma^\alpha_{xi}\sigma^\beta_{xi},
\label{terms}
\end{eqnarray}
where we have put $C={d-1 \over 2}$.

From Eqs.(\ref{Zn}), (\ref{U}) and (\ref{terms}),
\begin{equation}
[Z^n]=\exp \Big(-3\beta^2J^2N_P\sum_{\alpha<\beta}Q_{\alpha\beta}^4
-3J_0\beta N_P\sum_\alpha U^4_{0\alpha}
+N_L \log \mbox{Tr}\; e^L \Big),
\label{Zn2}
\end{equation}
where
\begin{equation}
L=4\beta^2J^2 C \sum_{\alpha<\beta}Q^3_{\alpha\beta}\sigma^\alpha_{xi}
\sigma^\beta_{xi}+ \beta \sum_\alpha (4J_0CU^3_{0\alpha}+h)\sigma^\alpha_{xi}.
\label{L}
\end{equation}
We assume a replica symmetric(RS) solution for $U_{0\alpha}=U_0$
and $Q_{\alpha\beta}=Q$.
In the RS case, $\log \mbox{Tr}\; e^L$ can be evaluated as follows,
\begin{eqnarray}
\log \mbox{Tr}\; e^L&=& \log \mbox{Tr}\; \sqrt{{4\beta^2J^2CQ^3 \over 2\pi}}
\int^\infty_{-\infty} dz \exp \Big(-{4\beta^2J^2CQ^3\over 2}z^2
+4\beta^2J^2CQ^3z\sum_\alpha
\sigma_{xi}^\alpha  \nonumber \\
&& \;    -2\beta^2J^2CnQ^3 
 +\beta (4J_0CU_0^3+h)\sum_\alpha\sigma_{xi}^\alpha
\Big) \nonumber  \\
&=& \log \Big(1+n\int^\infty_{-\infty} Dz \log (2\cosh \beta\tilde{H}(z))
-2n\beta^2J^2CQ^3+
O(n^2)\Big),
\label{log}
\end{eqnarray}
where
\begin{equation}
Dz=dz\; {e^{-{z^2\over 2}}\over \sqrt{2\pi}}, \;\; 
\tilde{H}(z)=2J\sqrt{CQ^3}z+4J_0CU^3_0+h.
\label{Dz}
\end{equation}
From (\ref{log}), the free energy $F_R$ is evaluated as
\begin{eqnarray}
-\beta F_R &=& \lim_{n\rightarrow 0}{[Z^n] -1 \over n} \nonumber \\
&=& N_P\Big(-{3\beta^2J^2 \over 2}(n-1)Q^4-3J_0\beta U_0^4
+{1\over Cn}\log \mbox{Tr}\; e^L\Big)  \nonumber  \\
&=& N_P\Big({3\over 2}\beta^2J^2Q^4-3J_0\beta U_0^4+
{1\over C}\int Dz\; \log (2\cosh \beta \tilde{H}(z))-2\beta^2
J^2Q^3\Big).
\label{FR}
\end{eqnarray}
The values of MF's $U_0$ and $Q$ are determined by the stationary condition
of $F_R$,
\begin{equation}
{\partial F_R \over \partial U_0}=0, \;\; 
{\partial F_R \over \partial Q}=0.
\label{UQeq}
\end{equation}
Numerical calculation is necessary for solving Eq.(\ref{UQeq}),
and the result is given in Fig.5.

Let us explain physical meanings of the ``order parameters"
$U_{0\alpha}$ and $Q_{\alpha\beta}$.
In the ordinary gauge theories with constant gauge coupling,
the confinement and deconfinement phases are distinguished
by the expectation value of the Wilson loop operator $W(C)$,
\begin{equation}
W(C)= \prod_{(xi)\in C}\sigma_{xi},
\label{wilson}
\end{equation}
where $C$ is a large closed loop on the original lattice.
If the system is in the confinement phase,
$\langle W(C) \rangle \propto e^{-\mbox{\footnotesize Area}(C)}$,
whereas in the deconfinement phase, 
$\langle W(C) \rangle \propto e^{-\mbox{\footnotesize Perimeter}(C)}$.

In the RGT, on the other hand, the ensemble average must be taken
in order to obtain physical quantities.
Then order parameter is given by $[\langle W(C) \rangle]$.
Nonvanishing of the MF $U_{0\alpha}$ means the perimeter law
$[\langle W(C) \rangle] \propto e^{-\mbox{\footnotesize Perimeter}(C)}$,
which indicates that the system is in the deconfinement phase, the
Higgs phase in the present case.
As shown in Fig.5, the RMFT predicts that the Higgs phase exists
in the RGT if the fluctuation of the RGC $J_p$ is not so large\cite{RGC}.
Result in Fig.5 also indicates the existence of 
a ``gauge glass" phase\cite{wang}.
In this phase, $U_{0\alpha}=0$ whereas $Q_{\alpha\beta}\neq 0$.
This means $[\langle W(C) \rangle] \propto e^{-\mbox{\footnotesize Area}(C)}$
whereas 
$[\langle W(C) \rangle^2] \propto e^{-\mbox{\footnotesize Perimeter}(C)}$.
This prediction of the gauge glass is itself interesting but the spatial
dimension must be probably large for its realization.\footnote{In this
paper we are considering systems only on the simple square or cubic lattice.
If we consider more complicated lattice or networks, effective spatial
dimensions increase and the gauge glass phase may appear.}

Reliability of the RMFT can be studied as in the usual spin glass
models like the Sherrington-Kirkpatrick model.
Parisi-type solutions for replica-symmetry breaking are also interesting.
These problems are under study and results will be reported in a future 
publication.


\section{Conclusion}

In this paper we explicitly showed the relationship between 
$Z_N$ gauge theories and Kitaev's model for quantum memory
and computations.
$Z_N$ gauge theories appear as a result of the spontaneous breakdown
of the $U(1)$ gauge theory with the ``Higgs field" of charge $N$.
The Higgs-phase limit of the $Z_N$ gauge systems corresponds to Kitaev's model.
Stability of Kitaev's model was discussed and it was shown that the errors
or noise represented by the term like $\sum X_{xi}$ induce the phase transition
to the confinement phase in the gauge-theory terminology.
In that phase, quantum memory and quantum computations are impossible.
Then we studied effects of the RGC which are also regarded as noise
and errors in quantum computations.
Static RGC gives no significant effect on the phase structure
whereas time-dependent RGC induces phase transitions including
that to the gauge-glass phase.
Application of the present studies to non-Abelian discrete gauge theory
is interesting and important for quantum computations.

\begin{center}
{\Large{\bf Acknowledgement}}
\end{center}
We acknowledge helpful discussions with T.Ohno and T.Matsui.


\newpage

\begin{figure}[H]
 \psfrag{higgs}{Higgs} \psfrag{confinement}{confinement}
 \psfrag{gauge}{$Z_N$ gauge theory}
 \psfrag{xymodel}{XY model} \psfrag{kappa}{$1/\kappa$}
 \psfrag{0}{0} \psfrag{g2}{$g^2$}
 \begin{center}
  \includegraphics[scale=0.9]{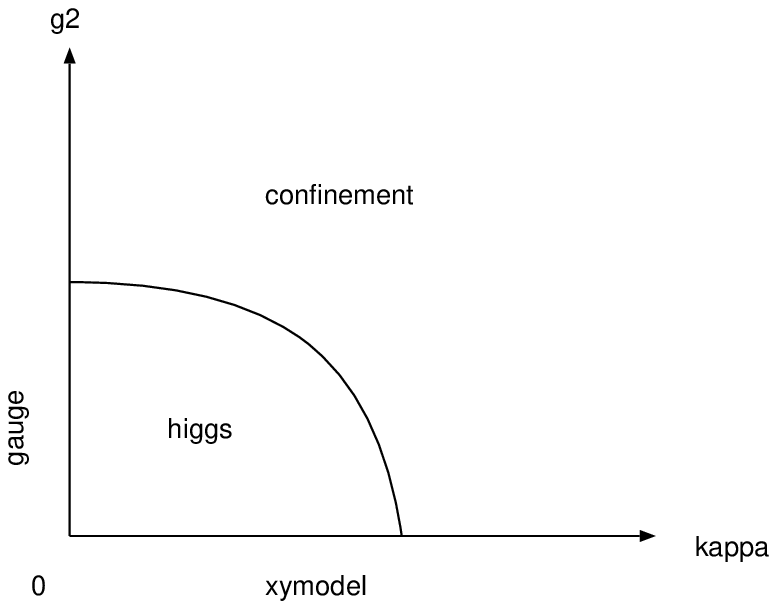}
  \caption{Phase diagram of the gauge-Higgs model}  
 \end{center}
\end{figure}

\vspace{100pt}

\begin{figure}[H]
 \psfrag{x}{$x^*$} \psfrag{y}{$y^*$}
 \begin{center}
  \includegraphics[scale=0.9]{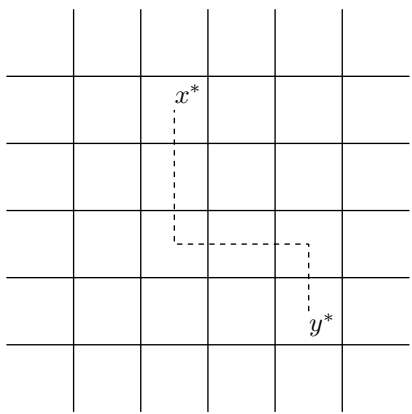}
  \caption{Path $\tilde{c}_{x^*y^*}$ connecting dual sites $x^*$ and $y^*$}
 \end{center}
\end{figure}

\begin{figure}[H]
 \begin{center}
  \begin{minipage}[b]{.32\linewidth}
   \begin{center}
    \psfrag{X}{$X_{x+1,2}$}\psfrag{x1}{$x+1$}
    \psfrag{Z}{$Z_{x1}$} \psfrag{x}{$x$}
    \includegraphics[scale=.4]{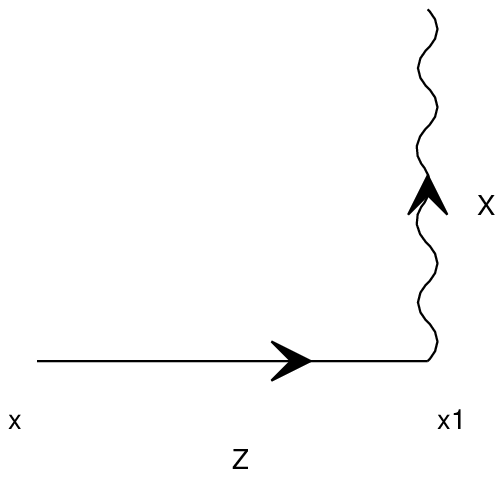}
   \end{center}
  \end{minipage}
    \begin{minipage}[b]{.32\linewidth}
   \begin{center}
    \psfrag{X2}{$X_{x+2,1}$}
    \psfrag{Z2}{$Z_{x2}$}
    \includegraphics[scale=.4]{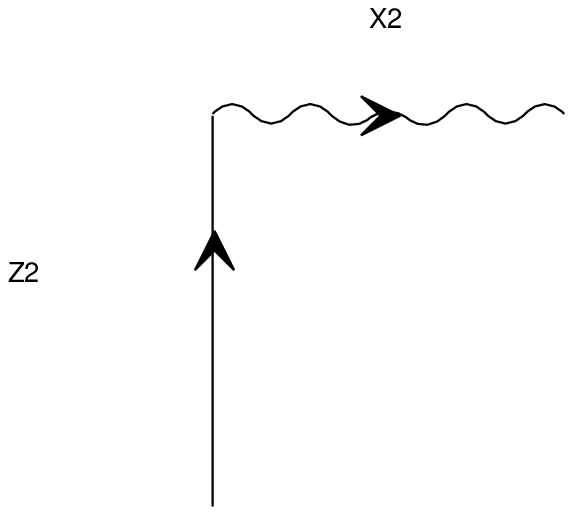}
   \end{center}
  \end{minipage}
  \begin{minipage}[b]{.32\linewidth}
   \begin{center}
    \psfrag{vortex}{vortex}
    \psfrag{charge}{charge}
    \includegraphics[scale=.6]{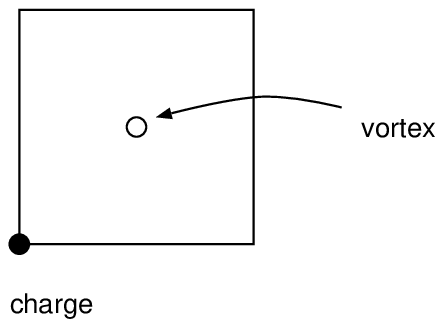}
   \end{center}
  \end{minipage}
 \end{center}
 \caption{The links $(\bar{x}i)$ on which $X_{\bar{x}i}^R$ operaters. 
  Dyon is composed of electric charge on site and magnetic vortex on
 plaquette}
\end{figure}

\vspace{30pt}

\begin{figure}[H]
 \begin{center}
  \psfrag{F}{\rotatebox[origin=c]{-90}{$\dfrac{F}{N_P}$}}
  \psfrag{U}{$U_0$}
  \psfrag{F(x,0.400)}{\footnotesize $\beta=0.4$}
  \psfrag{F(x,0.700)}{\footnotesize $\beta=0.7$}
  \psfrag{F(x,1.000)}{\footnotesize $\beta=1$}
  \psfrag{F(x,10.00)}{\footnotesize $\beta=10$}
  \includegraphics{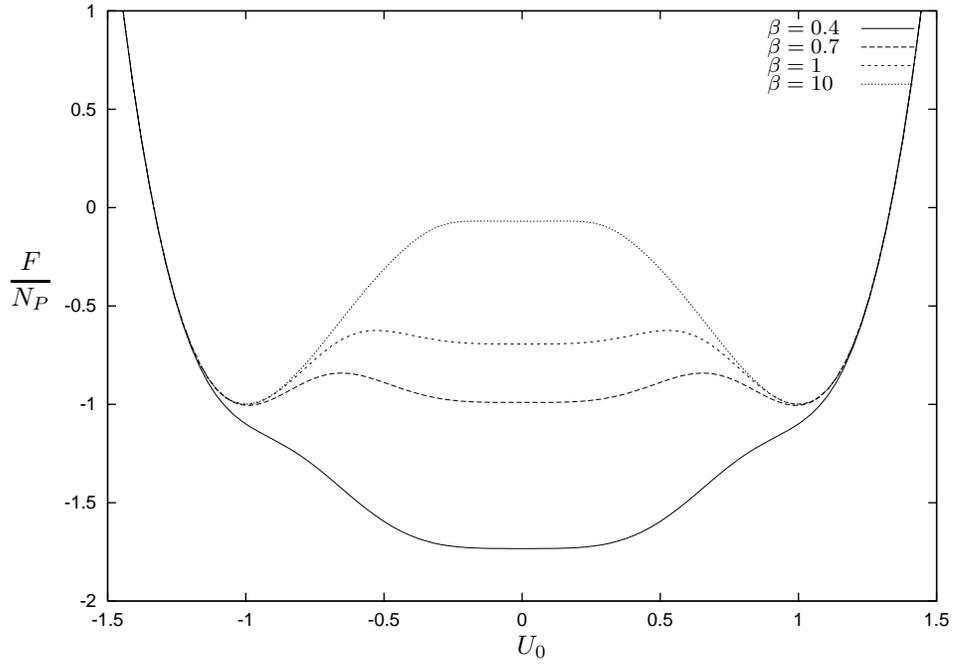}
 \end{center}
 \caption{Free energy of the $Z_2$ gauge system by the MFT}
\end{figure}

\begin{figure}[H]
 \begin{center}
  \psfrag{gauge}{gauge-glass}
  \psfrag{higgs}{Higgs}
  \psfrag{confinement}{confinement}
  \psfrag{0}{$0$}
  \psfrag{1.1}{$1.1$}
  \psfrag{0.5}{$0.5$}
  \psfrag{1/bj}{$k_B T/J$}
  \psfrag{j0/j}{$J_0/J$}
  \includegraphics[scale=0.9]{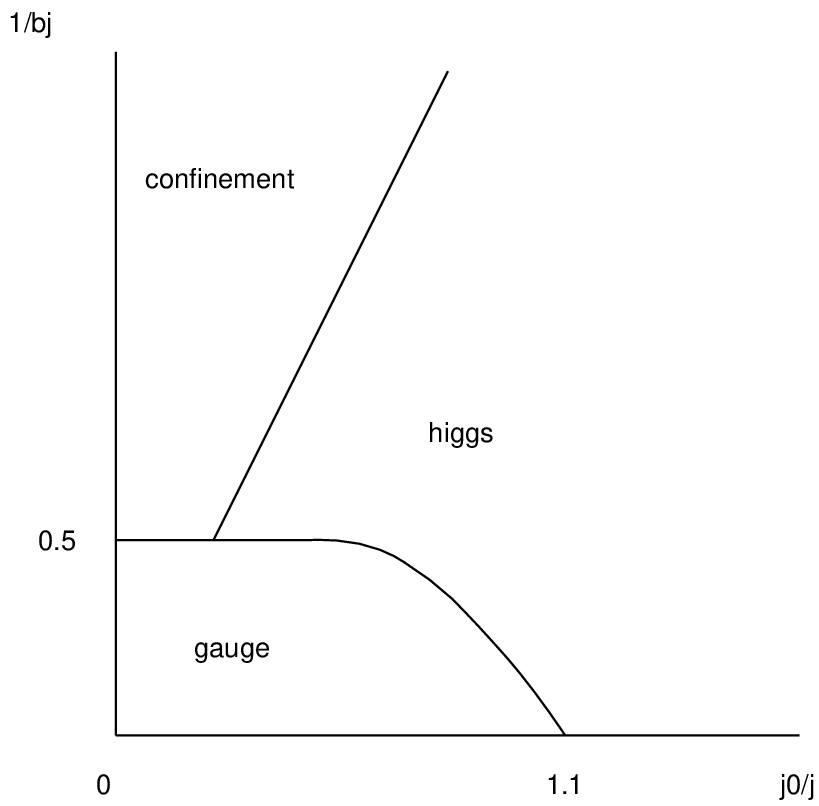}\\
  \vspace{50pt}
  \psfrag{gauge}{gauge-}
  \psfrag{glass}{glass}
  \psfrag{1/bj0}{$k_B T/J_0$}
  \psfrag{2}{$2$}

  \includegraphics[scale=0.9]{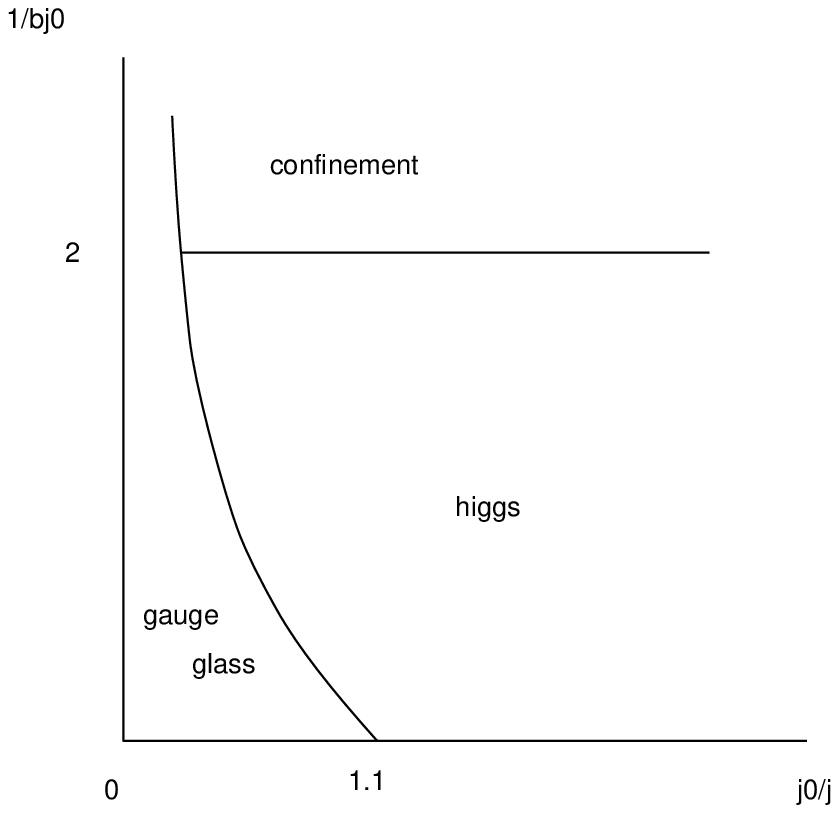}

  \caption{Phase diagram of the RGT obtained by the RMFT. The Higgs
  phase exists at low temperature(T) and small fluctuation of the
  RGC. Gauge-glass phase appears at low T and large fluctuation of the
  RGC. In the Higgs phase, the MF's $U_0 \neq 0$ and also $Q \neq 0$,
  whereas in the gauge-glass phase $U_0=0$ and $Q\neq 0$.}
 \end{center}
\end{figure}



\begin{thebibliography}{99}
\bibitem{wang2}Properties of the three-dimensional random $Z_2$ gauge theory 
at zero temperature are numerically investigated in Ref.\cite{wang}.
\bibitem{elitzur}As it is well-known, Elitzur's theorem forbids
the nonvanishing expectation value of the gauge-variant quantities
like $\sigma_{xi}$. Discussion on the compatibility of the MFT
and Elitzur's theorem is given in Ref.\cite{drouffe}.
\bibitem{RGC}Recent numerical studies on $3D$ $Z_2$ random gauge theory
in Refs.\cite{wang} and \cite{OAIM} indicate that the Higgs phase
at low temperature
disappears quite rapidly as the concentration $p$ of the ``wrong-sign"
plaquettes increases.
\end{thebibliography}
\end{document}